\documentstyle[12pt]{article}
\begin{document}
\begin{center}
{\Large Teleparallel Orthonormal Frames constraints on torsion detection?}
\vspace{1cm}
\noindent

L.C.Garcia de Andrade\footnote{Departamento de Fisica Teorica,Instituto de F\'{\i}sica , UERJ, Rua S\~{a}o
francisco Xavier 524, Rio de Janeiro,CEP:20550-013, Brasil.e-mail:garcia@dft.if.uerj.br.}
\end{center} 
\vspace{2cm}
\begin{center}
{\Large Abstract}
\end{center}
\vspace{0.5cm}
Spin-polarised cylindrically symmetric solution are shown not to be compatible with teleparallel gravity.This can be done in two distinct manners.The first is to show that not all components of the orthonormal tetrad (OT).It is argue however that this result maybe done by a bad choice of the spin distribution along the cylinder.A cylindrically symmetric Riemannian solution is obtained which represents a conical geometry of defects.   
\newpage
\section{Introduction}
Recently we showed that spin-polarised cylinders in the EC theory of gravity \cite{1,2,3} maybe used with the purposes of torsion detection \cite{4}.In the same paper we argue that spin polarised cylinders are not compatible with teleparallel gravity $T_{4}$ \cite{5}.Therefore teleparallel cylindrical geometry seems to be unpolarised or without spin density.Earlier Kopczy\'{n}ski \cite{6} and Nester \cite{7} engaged themselves on a discussion on the non-predictable behaviour of torsion in the experimentally viable one-parameter teleparallel theory of gravity.In their discussion examples were given where the presence of spin should be avoided as well.In a certain sense our example seems to corroborate this idea.We also make use of an orthonomal frame \cite{8} where the connection one-form components ${\omega}^{i}_{k}$ vanishes.By making the spin-density to vanish we obtain a Riemannian solution which represents a conical defect geometry \cite{9}.Let us consider here the Soleng \cite{3} geometry of cylinder in Riemann-Cartan geometry given by
\begin{equation}
ds^{2}=-(e^{\alpha}dt+Md{\phi})^{2}+r^{2}e^{-2{\alpha}}d{\phi}^{2}+e^{2{\beta}-2{\alpha}}(dr^{2}+dz^{2}).
\label{1}
\end{equation}
The functions ${\alpha},M$ and ${\beta}$ depend only on the coordinate r.An orthonormal comoving tetrad frame is defined by the basis forms 
\begin{equation} 
{\theta}^{0}=e^{\alpha}dt+Md{\phi} , 
\label{2} 
\end{equation}
\begin{equation}
{\theta}^{1}=e^{{\beta}-{\alpha}}dr ,
\label{3}
\end{equation}
\begin{equation}
{\theta}^{2}=re^{-{\alpha}}d{\phi} ,
\label{4}
\end{equation}
\begin{equation}
{\theta}^{3}=e^{{\beta}-{\alpha}}dz .
\label{5}
\end{equation}
Polarisation along the axis of symmetry is considered and the Cartan torsion is given in terms of differential forms by
\begin{equation}
T^{i}=2k{\sigma}{\delta}^{i}_{0}{\theta}^{1}{\wedge}{\theta}^{2}
\label{6}
\end{equation}
where ${\sigma}$ is a constant spin density.For computational convenience we addopt Soleng's definition \cite{3} for the RC rotation ${\Omega}$
\begin{equation}
{\Omega}=k{\sigma}+\frac{1}{2r}({\alpha}'M-{M'})e^{2{\alpha}-{\beta}}
\label{7}
\end{equation}
where ${\Omega}$ is the cylinder vorticity.Cartan's first structure equation is  
\begin{equation}
T^{i}=d{\theta}^{i}+{{\omega}^{i}}_{k}{\wedge}{\theta}^{k}
\label{8}
\end{equation}
and determines the connection forms ${{\omega}^{i}}_{j}$.The conection one-forms are given by
\begin{equation}
{\omega}^{0}_{1}={\alpha}'e^{{\alpha}-{\beta}}{\omega}^{0}-{\Omega}{\omega}^{2}
\label{9}
\end{equation}
\begin{equation}
{\omega}^{0}_{2}={\Omega}{\omega}^{1}
\label{10}
\end{equation}
\begin{equation}
{\omega}^{0}_{3}=0
\label{11}
\end{equation}
\begin{equation}
{\omega}^{1}_{2}=-{\Omega}{\omega}^{0}-(\frac{1}{r}-{\alpha}')e^{{\alpha}-{\beta}}{\omega}^{2}
\label{12}
\end{equation}
\begin{equation}
{\omega}^{1}_{2}=-({\beta}'-{\alpha}')e^{{\alpha}-{\beta}}{\omega}^{3}
\label{13}
\end{equation}
\begin{equation}
{\omega}^{2}_{3}=0
\label{14}
\end{equation}
From equation (\ref{10}) and the OT frame condition one obtains that vorticity vanishes,however imposing this constraint on the other equations we obtain an inconsistency ,namely that ${\alpha}$ is constant and that implies that ${\omega}^{1}_{2}$ does not vanish and therefore an OT frame cannot be found in this case and from the Cartan's second structure equation
\begin{equation}
{R^{i}}_{j}=d{{\omega}^{i}}_{j}+{{\omega}^{i}}_{k}{\wedge}{{\omega}^{k}}_{j}
\label{15}
\end{equation}
where the curvature RC forms ${R^{i}}_{j}={R^{i}}_{jkl}{\theta}^{k}{\wedge}{\theta}^{l}$ where ${R^{i}}_{jkl}$ is the RC curvature tensor, we note that the RC curvature tensor does not vanish,namely that the teleparallel condition is not obtained and therefore no teleparalelism is obtained from this OT frame condition with cylindrical symmetry.Nevertheless is possible to obtain teleparallel condition without OT frames since this OT frame condition are only sufficient from the mathematical point of view but not necessary.This is accomplished by computing the RC curvature components from the Cartan structure equations as
\begin{equation}
R_{0101}={\Omega}^{2}+[{\alpha}''+2{{\alpha}'}^{2}-{\alpha}'{\beta}']e^{2{\alpha}-2{\beta}} ,
\label{16}
\end{equation}
\begin{equation}
R_{0112}=[{\Omega}'+2{\alpha}'({\Omega}-{\sigma})]e^{{\alpha}-{\beta}} ,
\label{17}
\end{equation}
\begin{equation}
R_{0202}={\Omega}^{2}+[\frac{{\alpha}'}{r}-{({\alpha})'}^{2}]e^{2{\alpha}-2{\beta}} ,
\label{18}
\end{equation}
\begin{equation}
R_{0303}=[{\alpha}'{\beta}'-{{\alpha}'}^{2}]e^{2{\alpha}-2{\beta}} ,
\label{19}
\end{equation}
\begin{equation}
R_{0323}=-{\Omega}({\beta}'-{\alpha}')e^{{\alpha}-{\beta}} ,
\label{20}
\end{equation}
\begin{equation}
R_{1201}=[{\Omega}'+2{\Omega}{\alpha}']e^{{\alpha}-{\beta}} ,
\label{21}
\end{equation}
\begin{equation}
R_{1212}=3{\Omega}^{2}-2{\Omega}{\sigma}+({\alpha}''+\frac{{\beta}'}{r}-{\alpha}'{\beta}'+\frac{{\alpha}'}{r})e^{2{\alpha}-2{\beta}} ,
\label{22}
\end{equation}
\begin{equation}
R_{1313}=({\alpha}''-{\beta}'')e^{2{\alpha}-2{\beta}} ,
\label{23}
\end{equation}
\begin{equation}
R_{2303}=-{\Omega}({\beta}'-{\alpha}')e^{{\alpha}-{\beta}} ,
\label{24}
\end{equation}
\begin{equation}
R_{2323}=[\frac{{\alpha}'}{r}-\frac{{\beta}'}{r}+{\alpha}'{\beta}'-{{\alpha}'}^{2}]e^{2{\alpha}-2{\beta}} .
\label{25}
\end{equation}
We shall adopt here the simplest teleparallel condition ${R^{i}}_{jkl}=0$.When this $T_{4}$ constraint is applied to expressions (\ref{21}) and (\ref{21}) above simultaneously one is led to the constraint 
\begin{equation}
{\sigma}= 0
\label{26}
\end{equation}
which represents a spin unpolarised cylinder or without spin at all.Now a simple Riemannian  solution can be obtained by the vanishing of spin density ${\sigma}$ into the equation (\ref{7}) one obtains
\begin{equation}
{\Omega}=\frac{1}{2r}({\alpha}'M-{M'})e^{2{\alpha}-{\beta}}
\label{27}
\end{equation}
In the particular case when vorticity  vanishes ${\Omega}=0$ one obtains the following differential equation 
\begin{equation}
\frac{M'}{M}={\alpha}'
\label{28}
\end{equation}
where the dash represents the derivative with respect to the radial coordinate.By solving this equation one yields
\begin{equation}
M(r)=e^{\alpha}
\label{29}
\end{equation}
By substitution expression (\ref{29}) into the line element above we get
\begin{equation}
ds^{2}=-A(dt+d{\phi})^{2}+Cr^{2}d{\phi}^{2}+(dr^{2}+dz^{2})
\label{30}
\end{equation}
where A and C are constants with $C=A^{-1}$.This line element a Riemannian cosmic defect investigated previously by Letelier \cite{9} and Tod \cite{10} from the holonomy of connection.
\begin{flushleft}
{\large Acknowledgements}
\end{flushleft}
I would like to thank Professors J.Geraldo Pereira,H.Soleng and P.S.Letelier for helpful discussions on the subject of this paper.Financial support from CNPq. 
is gratefully acknowledged.


\newpage
\begin{thebibliography}{11}
\bibitem{1}L.C.Garcia de Andrade,Class. Quantum Gravity (2001) 18,in press.
\bibitem{2}M.L.Bedran and L.C.Garcia de Andrade,Prog.Theor.Phys.(1983)12,1583.
\bibitem{3}H.Soleng,Class. and Quant. Gravity 7,(1990),999.
\bibitem{4}C. L\"{a}mmerzahl,Phys.Lett. A 228 (1997)223.
\bibitem{5}A.Einstein and E.Cartan ,Letters on Absolute Parallelism (1979) Princeton University Press and A.Einstein,Matematische Annalen,(1930),685,Berlin.
\bibitem{6}W.Kopczy\'{n}ski,J.Phys.A:Math.Gen.15(1982)493.
\bibitem{7}J.Nester,Class. and Quantum Gravity 5(1988)1003.
\bibitem{8}J.Nester,J.Math.Phys. (1988).
\bibitem{9}P.S.Letelier,Class. Quantum Gravity (1995).
\bibitem{10}K.P.Tod,Class.Quantum and Grav.11(1994)1331.

\end{thebibliography}
\end{document}